\documentclass[seceq,supplement]{ptptex}

\usepackage{graphicx}
\usepackage{wrapft}

\def\bge{\begin{equation}}
\def\ene{\end{equation}}
\def\bg{\begin{eqnarray}}
\def\en{\end{eqnarray}}
\def\nn{\nonumber}

\def\ubar{\bar{u}}

\def\cbar{\bar{c}}
\def\qbar{\bar{q}}
\def\Qbar{\overline{Q}}
\def\Kbar{\overline{K}}
\def\Dbar{\overline{D}}
\def\D0bar{\overline{D^0}}
\def\Bbar{\overline{B}}

\title{
Properties of charmed and bottom hadrons in nuclear medium:\\
Results for $\Lambda_c^+$ and $\Lambda_b$ hypernuclei
}

\author{
K. Tsushima$^1$\footnote{Invited talk presented at YITP-RCNP Workshop on 
"Chiral Restoration in Nuclear Medium", October 7-9, 2002, YITP
Kyoto University, Japan, to be published in the proceedings.}, 
F.C. Khanna$^2$ 
}

\inst{
$^1$Department of Physics and Astronomy, 
University of Georgia, Athens, GA 30602, USA\\
$^2$Department of Physics, University of Alberta, Edmonton, Canada, T6G 2J1, 
and TRIUMF, 4004 Wesbrook Mall, Vancouver, B.C., V6T 2A3, Canada
}

\abst{
This reports our recent studies on 
changes in properties of heavy hadrons containing at least a charm or
a bottom quark in nuclear matter, 
and that the results for the $\Lambda^+_c$ and $\Lambda_b$ hypernuclei 
are studied quantitatively. 
Comparisons are made with the results for the $\Lambda$ hypernuclei 
studied previously in the same approach.
It is shown that 
although the scalar and vector potentials for the $\Lambda$,
$\Lambda_c^+$ and $\Lambda_b$ in the hypernuclei multiplet
with the same baryon numbers are quite similar,
the wave functions obtained, e.g., for $1s_{1/2}$ state, are very different.
The $\Lambda^+_c$ probability density distribution in $^{209}_{\Lambda^+_c}$Pb
is much more pushed away from the center
than that for the $\Lambda$ in $^{209}_\Lambda$Pb due to the Coulomb force.
On the contrary, the $\Lambda_b$ probability density distributions
in $\Lambda_b$ hypernuclei are much larger near the origin than those for
the $\Lambda$ in the $\Lambda$ hypernuclei 
due to its heavy mass.
A possibility of $B^-$ nuclear bound
(atomic) states is also briefly discussed.
}

\begin{document}

\maketitle

\section{Introduction}

Partial restoration of chiral symmetry
in nuclear medium is now one of the most 
important and interesting issues 
in nuclear and hadronic physics.
There is no doubt that it plays a crucial role in understanding  
numerous phenomenon involving many particles, such as 
relativistic heavy ion collisions, structure and properties 
of neutron stars, and those of heavy nuclei, and so on.   
Coverage of the issue is too vast  
to cite complete references.
This workshop is also one of the activities to understand and 
study the role of chiral restoration in nuclear medium.  

Our focus is on the consequences 
of partial restoration of chiral symmetry  
in nuclear medium, on the properties of heavy hadrons  
containing at least a charm or a bottom quark.
We report the results on the 
changes in properties of heavy hadrons 
in nuclear matter~\cite{matterbc}, and results 
for the $\Lambda_c^+$ and $\Lambda_b$ hypernuclei 
studied quantitatively~\cite{hypbc}.
In spite of the importance, there are studied 
for heavy mesons with charm only a limited number 
($J/\Psi$~\cite{Hayashigaki1,Klingl} and $D (\Dbar)$~\cite{Hayashigaki2}) 
in nuclear matter using the QCD sum rule. 
However, there seem to exist no studies for heavy baryons with a charm
or a bottom quark, except for the studies 
made recently~\cite{matterbc,hypbc}. 
Furthermore, in considering recent experiments on
high energy heavy ion collisions, to study general properties of heavy
hadrons in nuclear medium is essential, because
elementary hadronic reactions
occur in high nuclear density zone of the collisions, and
many hadrons produced there are under the 
influence of a surrounding nuclear medium.

Although the baryons with a charm or a bottom quark 
have a typical mean life of the order $10^{-12}$ seconds
(magnitude is shorter than hyperons), we would like to
gain an understanding of the movement of such a hadron in its
nucleonic environment.
The light quark in the hadron (and nucleons) would
change its property in nuclear medium in a self-consistent manner, and
will thus affect the overall interaction with nucleons.
With this understanding we will be in a better
position to learn about the hadron properties with the presence of heavy
quarks, or those for the hadrons containing 
heavy quarks in nuclear matter (in finite nuclei). 

The approved construction of the Japan Hadron Facility (JHF), 
with a beam energy of 50 GeV, will produce charmed hadrons
profusely and bottom hadrons in lesser
numbers but still with an intensity that is comparable
to the present hyperon production rates.
The production of charmonium ($\cbar c$), mesons with charm,
and baryons with charm quarks will be
sufficiently large to make it possible to
study charmed hypernuclei. 
In mid 70's, a possibility of such charmed hypernuclei
was predicted theoretically~\cite{Tyapkin,Dover}. There were also 
studies of possible experimental searches  
at the ARES facility~\cite{charmexp1},
and at the $c\tau$-factory~\cite{charmexp2}.
It is clear that the situation for the experiments to search for such
charmed and bottom hypernuclei is now becoming realistic and would be
realized at JHF.

      At JHF, in addition to charmed  and bottom hyperons, mesons with open
charm (bottom) like $D^- (\bar{c}d)$ ($B^- (\bar{u}b)$) will be
produced. Such mesons like $K^- (\bar{u}s)$ can form mesic atoms around finite
nuclei. The atomic orbits will be very
small and will thus probe the surface of light nuclei and will be within the
charge radii for heavier nuclei. Thus, at
least for light nuclei they will give a precise information about the charge
density.

To perform theoretical studies, at present we need to resort to a model which 
can describe the properties of finite nuclei as well as hadron properties
in nuclear medium based on the quark degrees of freedom.
With its simplicity and applicability, we use
quark-meson coupling (QMC) model~\cite{Guich}, which has been
extended and successfully applied to many
problems in nuclear
physics~\cite{Guichonf,Saitof,qmc2,qmcf,Blunden,Jin,qmcnuc}, 
hypernuclei~\cite{Tsushima_hyp}, and 
properties of hadrons in nuclear
medium~\cite{TsushimaK,qmcapp,Tsushimad}.
In particular, recent measurements of polarization transfer performed
at MAMI and Jlab~\cite{JlabQMC} support the medium modification of the
proton electromagnetic form factors calculated by the QMC model.
The final analysis~\cite{Strauch}
seems to become more in favor of QMC, although
still error bars may be too large to draw a definite conclusion.
This gives us some confidence in using QMC, and we hope 
it will provide us with a valuable glimpse into the
properties of charmed and bottom hypernuclei.

\section{Charmed and Bottom hypernuclei}

\subsection{Mean-field equations of motion}

We consider static, (approximately) spherically symmetric
charmed and bottom hypernuclei (closed shell plus one heavy baryon
configuration) ignoring small nonspherical
effects due to the embedded heavy baryon.
We adopt Hartree, mean-field approximation.
In this approximation, $\rho NN$ tensor coupling gives
a spin-orbit force for a nucleon bound
in a static spherical nucleus, although
in Hartree-Fock it can give a central force which contributes to
the bulk symmetry energy~\cite{Guichonf,Saitof}.
Furthermore, it gives no contribution for nuclear
matter since the meson fields are independent of position
and time. Thus, we ignore the $\rho NN$ tensor coupling
as usually adopted in the Hartree treatment of
quantum hadrodynamics (QHD)~\cite{QHD1,QHD2}.

Using the Born-Oppenheimer approximation, mean-field equations
of motion are derived for a charmed (bottom) hypernucleus
in which the quasi-particles moving
in single-particle orbits are three-quark clusters with the quantum numbers
of a charmed (bottom) baryon or a nucleon.
Then a relativistic
Lagrangian density at the hadronic
level~\cite{Guichonf,Saitof} can be constructed,
similar to that obtained in QHD~\cite{QHD1,QHD2},
which produces the same equations of motion
when expanded to the same order in velocity:
\begin{eqnarray}
{\cal L}^{CHY}_{QMC}=& &{\cal L}_{QMC} + {\cal L}^C_{QMC},
\label{LCHY}\\
{\cal L}_{QMC}=& &\overline{\psi}_N(\vec{r})
\left[ i \gamma \cdot \partial
- M_N^{\star}(\sigma) - (\, g_\omega \omega(\vec{r})
+ g_\rho \frac{\tau^N_3}{2} b(\vec{r})
+ \frac{e}{2} (1+\tau^N_3) A(\vec{r}) \,) \gamma_0
\right] \psi_N(\vec{r}) \quad \nn \\
-& &\frac{1}{2}[ (\nabla \sigma(\vec{r}))^2 +
m_{\sigma}^2 \sigma(\vec{r})^2 ]
+ \frac{1}{2}[ (\nabla \omega(\vec{r}))^2 + m_{\omega}^2
\omega(\vec{r})^2 ] \nn \\
+& &\frac{1}{2}[ (\nabla b(\vec{r}))^2 + m_{\rho}^2 b(\vec{r})^2 ]
+ \frac{1}{2} (\nabla A(\vec{r}))^2, 
\label{LQMC}\\
{\cal L}^C_{QMC}=& &\hspace{-1em}\sum_{C=\Lambda_c^+,\Lambda_b}
\hspace{-1em}\overline{\psi}_C(\vec{r})
\left[ i \gamma \cdot \partial
- M_C^{\star}(\sigma)
- (\, g^C_\omega \omega(\vec{r})
+ g^C_\rho I^C_3 b(\vec{r})
+ e Q_C A(\vec{r}) \,) \gamma_0
\right] \psi_C(\vec{r}), 
\label{LCQMC}
\end{eqnarray}
where $\psi_N(\vec{r})$ ($\psi_C(\vec{r})$)
and $b(\vec{r})$ are respectively the
nucleon (charmed and bottom baryon) and the $\rho$
meson (the time component in the third direction of
isospin) fields, while $m_\sigma$, $m_\omega$ and $m_{\rho}$ are
the masses of the $\sigma$, $\omega$ and $\rho$ meson fields.
$g_\omega$ and $g_{\rho}$ are the $\omega$-$N$ and $\rho$-$N$
coupling constants which are related to the corresponding
($u,d$)-quark-$\omega$, $g_\omega^q$, and
($u,d$)-quark-$\rho$, $g_\rho^q$, coupling constants as
$g_\omega = 3 g_\omega^q$ and
$g_\rho = g_\rho^q$~\cite{Guichonf,Saitof}.
Hereafter, we will use notations for the quark flavors,
$q \equiv u,d$ and $Q \equiv s,c,b$.
Note that in usual QMC (QMC-I)
the meson fields appearing
in Eqs.~(\ref{LQMC}) and~(\ref{LCQMC}) represent the quantum numbers and
Lorentz structure as those in QHD~\cite{QHD2},
corresponding, $\sigma \leftrightarrow \phi_0$,
$\omega \leftrightarrow V_0$ and $b \leftrightarrow b_0$,
and they are not directly connected with the
physical particles, nor quark model states.
Their masses in nuclear medium do not vary
in the present treatment.
For the other version of QMC (QMC-II), where masses of the meson fields
are also subject to the medium modification in a self-consistent
manner, see Ref.~\cite{qmc2}. However, for a proper
parameter set (set B) the typical results obtained
in QMC-II are very similar to those of QMC-I.
The difference is $\sim 16$ \% for the largest case, but
typically $\sim 10$ \% or less. (For the effective
masses of the hyperons, it is less than $\sim 8$ \%.)
In an approximation where the $\sigma$, $\omega$ and $\rho$ fields couple
only to the $u$ and $d$ quarks,
the coupling constants in the charmed (bottom) baryon
are obtained as $g^C_\omega = (n_q/3) g_\omega$, and
$g^C_\rho = g_\rho = g_\rho^q$, with $n_q$ being the total number of
valence $u$ and $d$ (light) quarks in the baryon $C$. $I^C_3$ and $Q_C$
are the third component of the baryon isospin operator and its electric
charge in units of the proton charge, $e$, respectively.
The field dependent $\sigma$-$N$ and $\sigma$-$C$
coupling strengths predicted by the QMC model,
$g_\sigma(\sigma)$ and  $g^C_\sigma(\sigma)$,
related to the Lagrangian densities,
Eqs.~(\ref{LQMC}) and~(\ref{LCQMC}), at the hadronic level are defined by:
\bg
M_N^{\star}(\sigma) &\equiv& M_N - g_\sigma(\sigma)
\sigma(\vec{r}) ,  \\
M_C^{\star}(\sigma) &\equiv& M_C - g^C_\sigma(\sigma)
\sigma(\vec{r}) , \label{coupny}
\en
where $M_N$ ($M_C$) is the free nucleon (charmed and bottom baryon)
mass (masses).
Note that the dependence of these coupling strengths on the applied
scalar field must be calculated self-consistently within the quark
model~\cite{Guichonf,Saitof,Tsushima_hyp}.
Hence, unlike QHD~\cite{QHD1,QHD2,ruf,mar,Jennings}, even though
$g^C_\sigma(\sigma) / g_\sigma(\sigma)$ may be
2/3 or 1/3 depending on the number of light quarks in the baryon
in free space ($\sigma = 0$)\footnote{Strictly, this is true
only when the bag radii of nucleon and baryon $C$ are exactly the same
in the present model. See Eq.~(\ref{CC}).},
this will not necessarily  be the case in
nuclear matter.
More explicit expressions for $g^C_\sigma(\sigma)$
and $g_\sigma(\sigma)$ will be given later.
From the Lagrangian density,
Eq.~(\ref{LCHY}), a set of
equations of motion for the charm
or bottom hypernuclear system is obtained,
\begin{eqnarray}
& &[i\gamma \cdot \partial -M^\star_N(\sigma)-
(\, g_\omega \omega(\vec{r}) + g_\rho \frac{\tau^N_3}{2} b(\vec{r})
 + \frac{e}{2} (1+\tau^N_3) A(\vec{r}) \,)
\gamma_0 ] \psi_N(\vec{r}) = 0, \label{eqdiracn}\\
& &[i\gamma \cdot \partial - M^\star_C(\sigma)-
(\, g^C_\omega \omega(\vec{r}) + g_\rho I^C_3 b(\vec{r})
+ e Q_C A(\vec{r}) \,)
\gamma_0 ] \psi_C(\vec{r}) = 0, \label{eqdiracy}\\
& &(-\nabla^2_r+m^2_\sigma)\sigma(\vec{r}) =
- [\frac{\partial M_N^\star(\sigma)}{\partial \sigma}]\rho_s(\vec{r})
- [\frac{\partial M_C^\star(\sigma)}{\partial \sigma}]\rho^C_s(\vec{r}),
\nn \\
& & \hspace{7.5em} \equiv g_\sigma C_N(\sigma) \rho_s(\vec{r})
    + g^C_\sigma C_C(\sigma) \rho^C_s(\vec{r}) , \label{eqsigma}\\
& &(-\nabla^2_r+m^2_\omega) \omega(\vec{r}) =
g_\omega \rho_B(\vec{r}) + g^C_\omega
\rho^C_B(\vec{r}) ,\label{eqomega}\\
& &(-\nabla^2_r+m^2_\rho) b(\vec{r}) =
\frac{g_\rho}{2}\rho_3(\vec{r}) + g^C_\rho I^C_3 \rho^C_B(\vec{r}),
 \label{eqrho}\\
& &(-\nabla^2_r) A(\vec{r}) =
e \rho_p(\vec{r})
+ e Q_C \rho^C_B(\vec{r}) ,\label{eqcoulomb}
\end{eqnarray}
where, $\rho_s(\vec{r})$ ($\rho^C_s(\vec{r})$), $\rho_B(\vec{r})$
($\rho^C_B(\vec{r})$), $\rho_3(\vec{r})$ and
$\rho_p(\vec{r})$ are the scalar, baryon, third component of isovector,
and proton densities at the position $\vec{r}$ in charmed or bottom 
hypernucleus~\cite{Guichonf,Saitof,Tsushima_hyp}.
On the right hand side of Eq.~(\ref{eqsigma}),
$- [\frac{\partial M_N^\star(\sigma)}{\partial \sigma}] =
g_\sigma C_N(\sigma)$ and
$- [\frac{\partial M_C^\star(\sigma)}{\partial \sigma}] =
g^C_\sigma C_C(\sigma)$, where $g_\sigma \equiv g_\sigma (\sigma=0)$ and
$g^C_\sigma \equiv g^C_\sigma (\sigma=0)$,
are a new, and characteristic feature of QMC
beyond QHD~\cite{QHD1,QHD2,ruf,mar,Jennings}.
The effective mass for the charmed or bottom baryon $C$ is defined by,
\begin{equation}
\frac{\partial M_C^\star(\sigma)}{\partial \sigma}
= - n_q g_{\sigma}^q \int_{bag} d\vec{x}
\ {\overline \psi}_q(\vec{x}) \psi_q(\vec{x})
\equiv - n_q g_{\sigma}^q S_C(\sigma) = - \frac{\partial}{\partial \sigma}
\left[ g^C_\sigma(\sigma) \sigma \right],
\label{gsigma}
\end{equation}
with the MIT bag model quantities, and
the in-medium bag radius satisfying the
mass stability condition~\cite{Guich,Guichonf,Saitof,Tsushima_hyp}:
\begin{eqnarray}
& &M_C^\star(\sigma) =
\sum_{j=q,Q}\frac{n_j\Omega^*_j -  z_C}{R_C^*}
+ \frac{4}{3}\pi ({R_C^*})^3 B,
\label{MC}\\
& &S_C(\sigma) = 
\left[\Omega_q^*/2+m_q^*R_C^*(\Omega_q^*-1)\right]/
\left[\Omega_q^*(\Omega_q^*-1)+ m_q^*R_C^*/2\right],
\label{SC}\\
& &
\Omega_q^* = \sqrt{x_q^2 + (R_C^* m_q^*)^2},\,\, 
\Omega_Q^* = \sqrt{x_Q^2 + (R_C^* m_Q)^2},\,\, 
m_q^* = m_q - g_{\sigma}^q \sigma (\vec{r}), 
\label{Omega}\\
& &C_C(\sigma) = \frac{S_C(\sigma)}{S_C(0)},\quad 
g^C_{\sigma} \equiv n_q g_{\sigma}^q S_C(0)
= \frac{n_q}{3} g_\sigma \frac{S_C(0)}{S_N(0)}
\equiv \frac{n_q}{3}g_\sigma \Gamma_{C/N}, 
\label{CC}\\
& & \left. dM^\star_{C}/dR_{C}
\right|_{R_{C} = R^*_{C}} = 0.
\label{bagradii}
\end{eqnarray}
Quantities for the nucleon are similarly obtained by replacing the
indices, $C \to N$.
Here, the MIT bag model quantities are calculated in a local
density approximation using the
spin and spatial part of the wave functions,
$\psi_f (x) = N_f e^{- i \epsilon_f t / R_h^*}\psi_f (\vec{x})$
($N_f$: the normalization factor), where 
the wave functions, $\psi_f (x)$, satisfy the Dirac equations for
the flavor $f$ quarks (and antiquarks) in the hadron bag
centered at a position $\vec{r}$ of the nucleus,
approximating the constant, mean, meson fields
within the bag (and neglecting the Coulomb force)
($|\vec{x} - \vec{r}|\le$
bag radius~\cite{TsushimaK,Tsushimad}):
\begin{eqnarray}
\left[ i \gamma \cdot \partial_x -
(m_q - V^q_\sigma(\vec{r})
\mp \gamma^0
\left( V^q_\omega(\vec{r})) +
\frac{1}{2} V^q_\rho(\vec{r})
\right) \right]
\left( \begin{array}{c} \psi_u(x) \\
\psi_{\bar{u}}(x) \\ \end{array} \right) &=& 0,
\label{Diracu}\\
\left[ i \gamma \cdot \partial_x -
(m_q - V^q_\sigma(\vec{r}))
\mp \gamma^0
\left( V^q_\omega(\vec{r}) -
\frac{1}{2} V^q_\rho(\vec{r})
\right) \right]
\left( \begin{array}{c} \psi_d(x) \\
\psi_{\bar{d}}(x) \\ \end{array} \right) &=& 0,
\label{Diracd}\\
\left[ i \gamma \cdot \partial_x - m_{Q} \right]
\psi_{Q} (x)\,\, ({\rm or}\,\, \psi_{\Qbar}(x)) &=& 0.
\label{DiracQ}
\end{eqnarray}
The (constant) mean-field potentials within the bag centered at
the position $\vec{r}$ of the nucleus,
are defined by $V^q_\sigma(\vec{r}) \equiv g^q_\sigma \sigma(\vec{r})$,
$V^q_\omega(\vec{r}) \equiv g^q_\omega \omega(\vec{r})$ and
$V^q_\rho(\vec{r}) \equiv g^q_\rho b(\vec{r})$,
with $g^q_\sigma$, $g^q_\omega$ and
$g^q_\rho$ the corresponding quark-meson coupling constants.
%
%
In Eqs.~(\ref{MC})~-~(\ref{bagradii}) 
$z_C$, $B$, $x_{q,Q}$, and $m_{q,Q}$ are the parameters
for the sum of the c.m.
and gluon fluctuation effects,
bag pressure, lowest eigenvalues for the quarks, $q$ or $Q$, respectively,
and the corresponding current quark masses.
$z_N$ and $B$ ($z_C$) are fixed by fitting the nucleon
(charmed or bottom baryon) mass
in free space. The current quark masses for the quarks, we use,
$(m_{u,d},m_s,m_c,m_b) = (5,250,1300,4200)$ MeV,
and $B = (170.0$ MeV$)^4$ is obtained by choosing the
bag radius for the nucleon in free space, $R_N = 0.8$ fm.
Calculated bag radii in free space, and the bag parameters
are $(R_\Lambda, R_{\Lambda^+_c}, R_{\Lambda_b})$
$= (0.806, 0.846, 0.930)$ fm, and 
$(z_N,\, z_\Lambda,z_{\Lambda^+_c},z_{\Lambda_b})$
$= (3.295, 3.131, 1.766, -0.643)$, respectively.
The parameters associated with the $u$, $d$ and $s$ quarks are the same as in
the previous studies~\cite{Saitof,Tsushima_hyp}.
The parameters at the hadron level, which are already fixed by the study of
nuclear matter and finite nuclei~\cite{Saitof},
are as follows: $m_\omega = 783$ MeV, $m_\rho = 770$ MeV, $m_\sigma = 418$ MeV,
$e^2/4\pi = 1/137.036$, $g^2_\sigma/4\pi = 3.12$, $g^2_\omega/4\pi = 5.31$
and $g^2_\rho/4\pi = 6.93$.
Concerning the sign of $m_q^*$ in (hyper)nucleus in Eq.~(\ref{Omega}),
it reflects nothing but the strength
of the attractive (negative) scalar potential, 
and thus naive interpretation of the mass for a physical particle,
which is positive, should not be applied.

At the hadronic level, the entire information
on the quark dynamics is condensed into the effective coupling
$C_{N,C}(\sigma)$ of Eq.~(\ref{eqsigma}).
Furthermore, when $C_{N,C}(\sigma) = 1$, which corresponds to
a structureless nucleon or heavy baryon $C$, the equations of motion
given by Eqs.~(\ref{eqdiracn})-(\ref{eqcoulomb})
can be identified with those derived
from QHD~\cite{mar,ruf,Jennings},
except for the terms arising from the tensor coupling and the non-linear
scalar field interaction introduced beyond naive QHD.

%
\subsection{Nuclear matter limit}

Here, we consider a charmed or a bottom hadron 
in nuclear matter. In this limit the meson fields
become constant, and we denote the mean-value of the $\sigma$ field as
$\overline{\sigma}$.
Furthermore, under this
limit, we can also generally consider a hadron, $h$, 
embedded in the nuclear matter 
in the same way as that for the charmed (bottom) baryon.
(For example, a Lagrangian density for a meson-nuclear 
system can be also written in a similar way, 
if ${\cal L}^C_{QMC}$ is replaced by the corresponding meson Lagrangian
density in Eq.(\ref{LCQMC}).)
The self-consistency condition for the $\sigma$ field, $\overline{\sigma}$,
is given by~\cite{Guich,Guichonf,Saitof}, 
\bg
\overline{\sigma}&=&\frac{g_\sigma }{m_\sigma^2}C_N(\overline{\sigma})
\frac{4}{(2\pi)^3}\int d\vec{k} \theta (k_F - k)
\frac{M_N^{\star}(\overline{\sigma})}
{\sqrt{M_N^{\star 2}(\overline{\sigma})+\vec{k}^2}} , \label{scc}
\en
where $g_{\sigma} = (3 g^q_\sigma S_N(0))$, $k_F$ is the
Fermi momentum,
and $C_N(\overline{\sigma})$ is now the constant value of $C_N$ in the
scalar field.
Note that $M_N^\star (\overline{\sigma})$,
in Eq.~(\ref{scc}), must be calculated
self-consistently by the MIT bag model, through
Eqs.~(\ref{gsigma})~-~(\ref{bagradii}).
This self-consistency equation for $\overline{\sigma}$
is the same as that in QHD, except that in the latter model one has
$C_N(\overline{\sigma})=1$~\cite{QHD1,QHD2}.
Using the obtained mean field value, $\overline{\sigma}$,
the corresponding quantity for the hadron $h$ in nuclear matter  
can be also calculated using Eqs.~(\ref{gsigma})~-~(\ref{bagradii}),
where the effect of a single hadron on the 
mean field value, $\overline{\sigma}$, in nuclear
matter can be neglected.

\begin{figure}[htb]
\parbox{\halftext}{
\includegraphics[width=5.5cm,angle=-90]{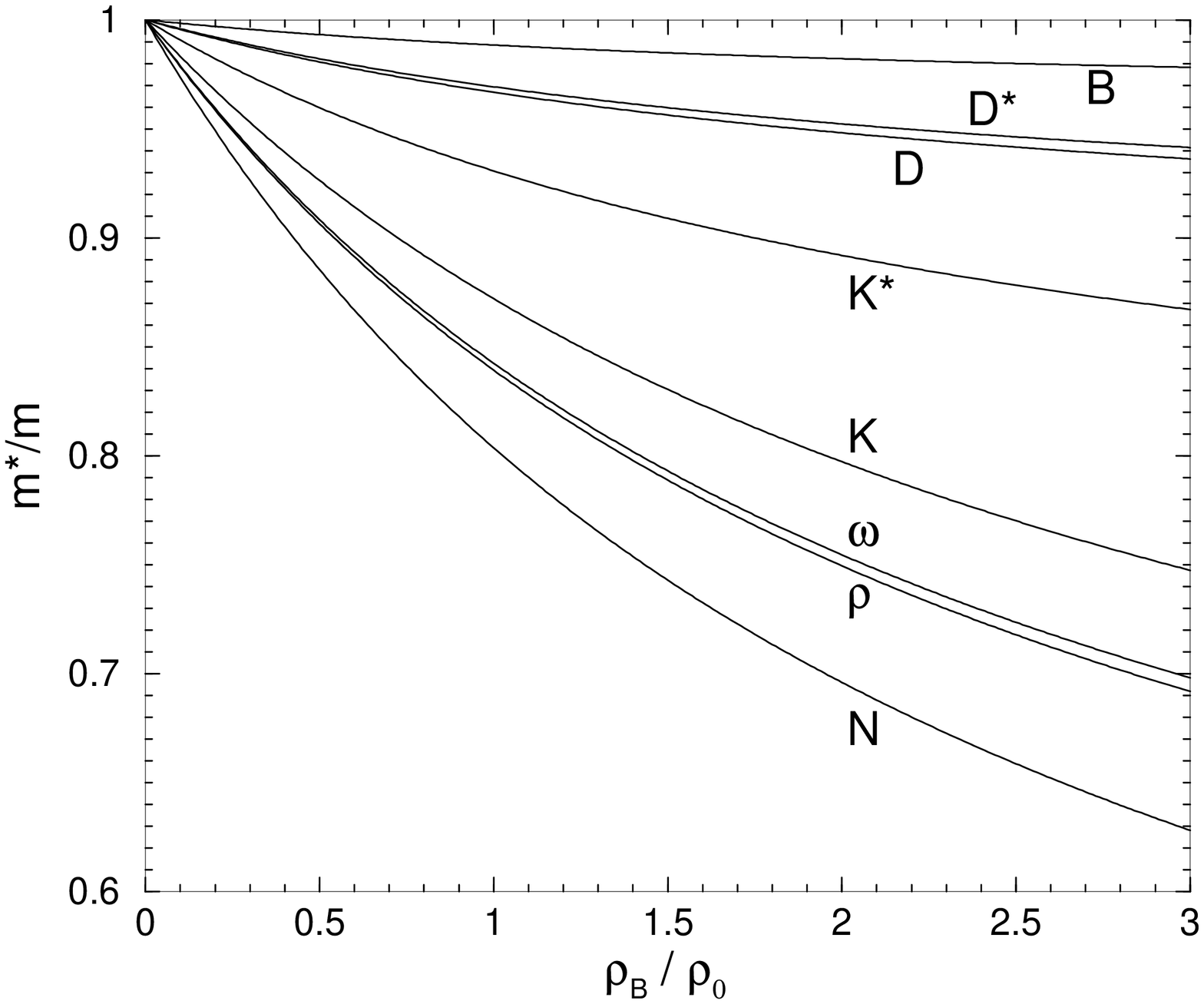}
\caption{Effective mass ratios for mesons in nuclear matter, where,
$\rho_0 = 0.15$ fm$^{-3}$.
}
\label{mesonmass}
}
\hfill 
\parbox{\halftext}{
\vspace{-1em}
\includegraphics[width=5.5cm,angle=-90]{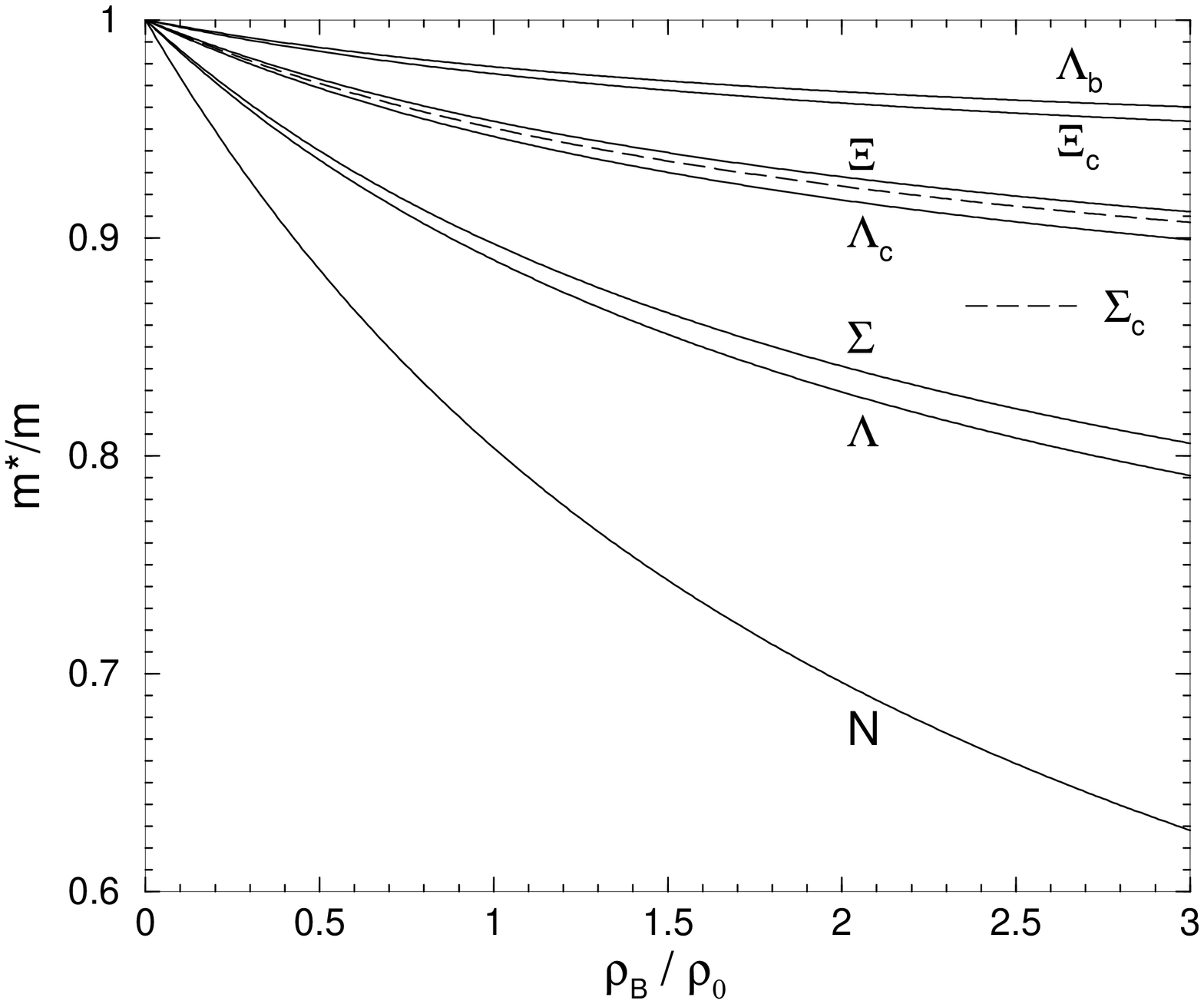}
\caption{Effective mass ratios for baryons. 
}
\label{baryonmass}
}
\end{figure}

In Figs.~\ref{mesonmass} and~\ref{baryonmass} we show the calculated
ratios of effective masses versus those of the free.
With increasing density the ratios decrease as usually expected,
but decrease in magnitude is from larger to smaller:
hadrons with only light
quarks, with one strange quark, with one charm quark, and with one
bottom quark. This is because their masses in free space
are in the order from light to heavy. Thus, the net ratios for the
decrease in masses (developing of scalar masses) compared to that of
the free masses becomes smaller.
This may be regarded as a measure of the role of light
quarks in each hadron system in nuclear matter, in a sense by how
much do they lead to a partial restoration of the chiral
symmetry.

\begin{wrapfigure}{l}{6.5cm}
\includegraphics[width=5.5cm,angle=-90]{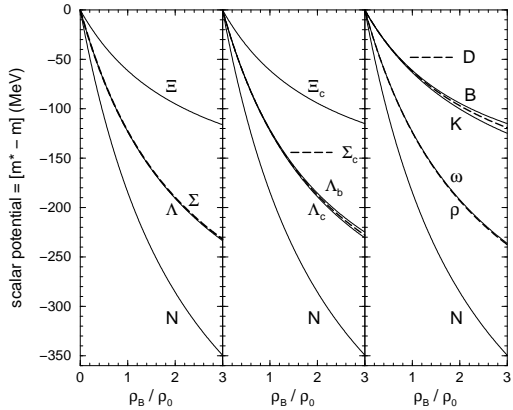}
\caption{Scalar potentials for various hadrons.}
\label{spotential}
\end{wrapfigure}

We next compare the 
scalar potentials of hadrons in nuclear matter.
The scalar, $V^{h}_s$, and vector, $V^{h}_v$, potentials
for the hadrons $h$,
in nuclear matter are given by,
\bg
V^h_s &=& m^*_h - m_h,\,\,\label{spot}
\\
V^h_v &=&
  (n_q - n_{\bar{q}}) {V}^q_\omega + I^h_3 V^q_\rho,
\qquad (V^q_\omega \to 1.4^2 {V}^q_\omega\,\,
{\rm for}\, K,\Kbar,D,\Dbar,B,\Bbar),
\label{vpot}
\en
where $I^h_3$ is the third component of isospin projection
of the hadron $h$, and $n_q$ ($n_{\qbar}$) is the number of 
light quarks (antiquarks) in
the hadron $h$. Thus, the vector potential for a heavy baryon
containing at least a charm or bottom quark, 
is equal to that of the hyperon with
the same light quark configuration in QMC.
Note that, in studies of the kaon system, we found that it was
phenomenologically necessary to increase the strength of the vector
coupling to the non-strange quarks in the $K^+$,  
i.e., $g_{K\omega}^q \equiv 1.4^2 g^q_\omega$, 
to reproduce the empirically extracted $K^+$-nucleus
interaction~\cite{TsushimaK}. This may be related to the fact that
kaon is a pseudo-Goldstone boson, where treatment of the Goldstone
bosons in a naive quark model is usually unsatisfactory.
We assume this, $g^q_\omega \to 1.4^2 g^q_\omega$,
also for the $D$, $\Dbar$~\cite{Tsushimad},
$B$ and $\Bbar$ to allow an upper limit situation.
Calculated scalar potentials 
are shown in Fig.~\ref{spotential}.
From the results it is confirmed that
the scalar potential for the hadron $h$, $V_s^h$,
follows a simple light quark number scaling rule:
\bge
V_s^h \simeq \left[(n_q + n_{\qbar}) V_s^N\right] / 3,
\ene
where $V_s^N$ is the scalar potential for the nucleon.
It is interesting to notice that, the baryons with charm and bottom quarks
($\Xi_c=(qsc)$), show very similar
features to those of the corresponding hyperons with strange quarks.
Then, we can naively expect at this stage that these heavy baryons 
will also form charmed (bottom) hypernuclei, as the
hyperons with strangeness do. (Recall that the repulsive, vector
potentials are the same for the corresponding hyperons with the same light
quark configurations.)

In addition, $B^-$ meson will also certainly form meson-nuclear bound states,
because $B^-$ meson is $\ubar b$ and feels a strong attractive
vector potential in addition to the attractive 
Coulomb force. This makes it much easier to be bound in a
nucleus compared to the $D^0$~\cite{Tsushimad}, which is $c \ubar$
and blind to the Coulomb 
force.  
This reminds us of a situation
for the kaonic ($K^- (\ubar s)$) atom~\cite{katom1,katom2}.
Such atoms like $B^- (\ubar b)$ atoms will have the
meson much closer to the nucleus and will thus probe
even smaller changes in the
nuclear density. This will be a complementary information to
the $D^- (\bar{c} d)$ nuclear bound states, which would provide us 
details of the vector potential in a nucleus~\cite{Tsushimad}.
%

\section{Results for $\Lambda_c^+$ and $\Lambda_b$ hypernuclei}

Here, we present results for the $\Lambda_c^+$ 
and $\Lambda_b$ hypernuclei, and compare them with those
for the $\Lambda$ hypernuclei studied 
previously~\cite{Tsushima_hyp} in QMC.

We briefly discuss the spin-orbit force in QMC~\cite{Guichonf}.
(See Refs.~\cite{Guichonf} and~\cite{Tsushima_hyp} for detail.)
To include the spin-orbit potential approximately correctly, 
e.g., for the $\Lambda^+_c$, 
we add perturbatively the correction,
$ -\frac{2}{2 M^{\star 2}_{\Lambda^+_c} (\vec{r}) r}
\, \left( \frac{d}{dr} g^{\Lambda^+_c}_\omega \omega(\vec{r}) \right)
\vec{l}\cdot\vec{s}$, 
to the single-particle energies obtained with the Dirac
equation~\cite{Tsushima_hyp}. 
This may correspond to a correct spin-orbit force which
is calculated by the underlying quark
model~\cite{Guichonf,Tsushima_hyp}:
\begin{equation}
V^{\Lambda^+_c}_{S.O.}(\vec{r}) \vec{l}\cdot\vec{s}
= - \frac{1}{2 M^{\star 2}_{\Lambda^+_c} (\vec{r}) r}
\, \left( \frac{d}{dr} [ M^\star_{\Lambda^+_c} (\vec{r})
+ g^{\Lambda^+_c}_\omega \omega(\vec{r}) ] \right) \vec{l}\cdot\vec{s},
\label{soQMC}
\end{equation}
since the Dirac equation at the hadronic level in usual QHD-type
models leads to:
\begin{equation}
V^{\Lambda^+_c}_{S.O.}(\vec{r}) \vec{l}\cdot\vec{s}
= - \frac{1}{2 M^{\star 2}_{\Lambda^+_c} (\vec{r}) r}
\, \left( \frac{d}{dr} [ M^\star_{\Lambda^+_c} (\vec{r})
- g^{\Lambda^+_c}_\omega \omega(\vec{r}) ] \right) \vec{l}\cdot\vec{s},
\label{soQHD}
\end{equation}
which has the opposite sign for the vector potential,
$g^{\Lambda^+_c}_\omega \omega(\vec{r})$.
The correction to the spin-orbit force, which appears naturally in the
QMC model, may also be modeled at the hadronic level of the Dirac equation by
adding a tensor interaction, motivated by the quark
model~\cite{Jennings2,Cohen}. 

Here, we should make a comment that, as was discussed
by Dover and Gal~\cite{Gal}
in detail, one boson exchange model with underlying (approximate)
SU(3) symmetry in strong interaction, also leads to the weaker
spin-orbit forces for the (strange) hyperon-nucleon ($YN$)
than that for the nucleon-nucleon ($NN$).

\begin{wrapfigure}{l}{6.0cm}
\includegraphics[width=6.0cm,height=6.0cm,angle=-90]{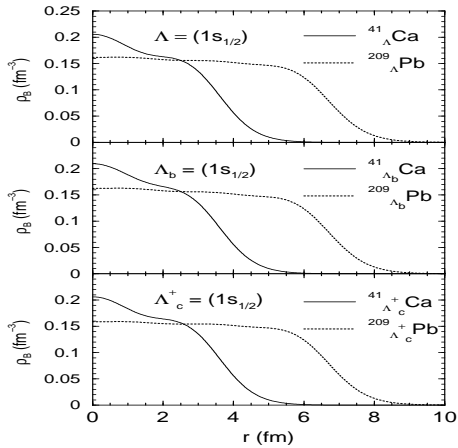}
\caption{Total baryon density distributions in
$^{41}_j$Ca and $^{209}_j$Pb ($j = \Lambda,\Lambda^+_c,\Lambda_b$),
for $1s_{1/2}$ state for the $\Lambda, \Lambda^+_c$ and $\Lambda_b$.
}
\label{CaPbden}
\end{wrapfigure}

However, in practice, because of its heavy mass
($M^\star_{\Lambda^+_c}$), contribution to the single-particle energies
from the spin-orbit potential, with or without
including the correction term, turned
out to be even smaller than that for the
$\Lambda$ hypernuclei, and further 
smaller for the $\Lambda_b$ hypernuclei~\cite{hypbc}.

First, we show in Fig.~\ref{CaPbden} the total baryon density distributions
in $^{41}_j$Ca and $^{209}_j$Pb ($j=\Lambda,\Lambda^+_c,\Lambda_b$),
for $1s_{1/2}$ state in each hypernucleus.
Note that because of the self-consistency, the total baryon density
distributions are dependent on the state of
the embedded particles. The total baryon density distributions 
obtained are quite
similar for the $\Lambda$, $\Lambda^+_c$ and $\Lambda_b$ hypernuclei
multiplet with the same baryon numbers, A, since the effect of
$\Lambda,\Lambda^+_c$ and $\Lambda_b$ is $\cong 1/A$.

\begin{figure}[htb]
\parbox{\halftext}{
\includegraphics[width=6.0cm,height=6.0cm,angle=-90]{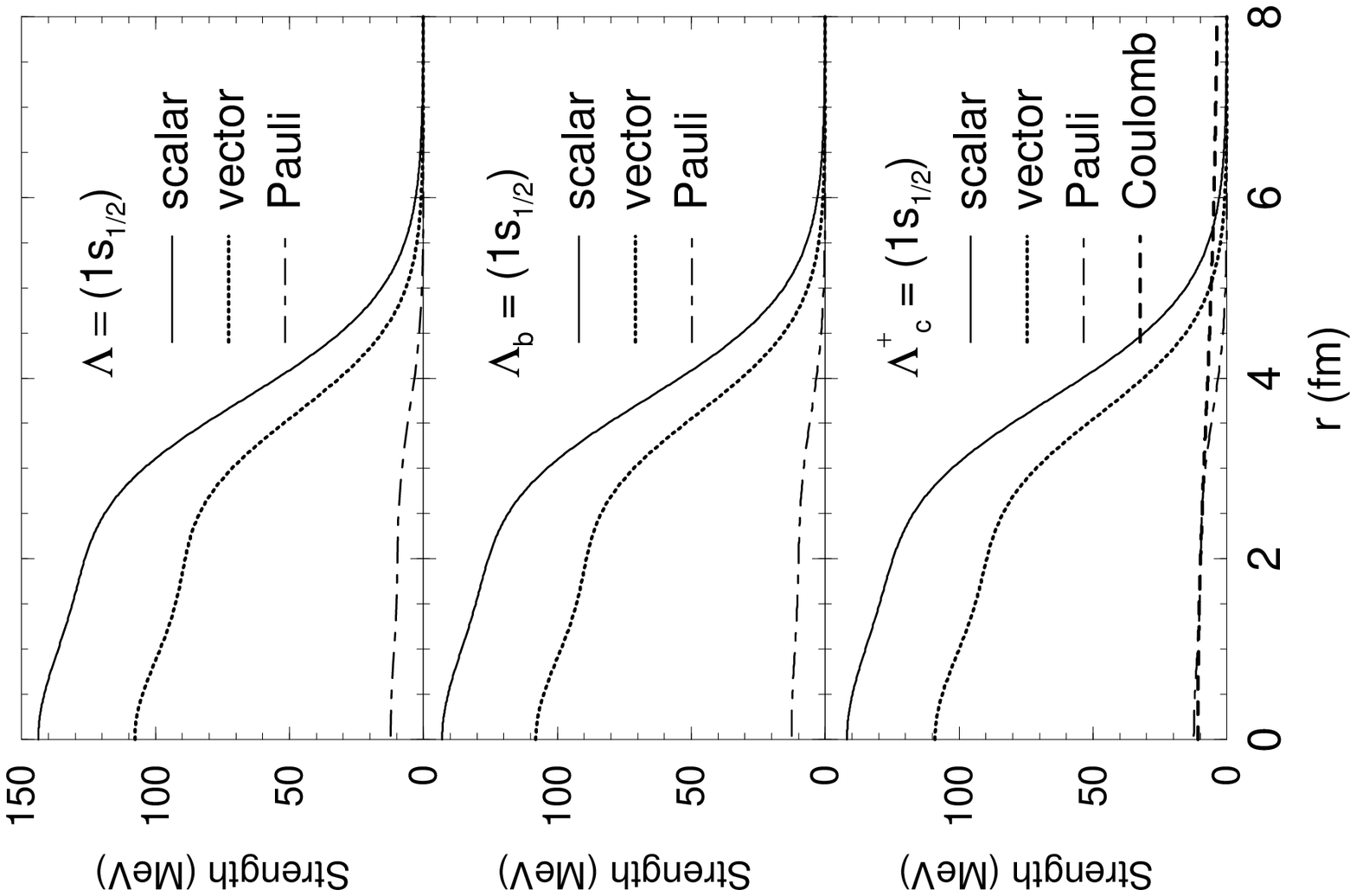}
\caption{Potential strengths for $1s_{1/2}$ state for 
the $\Lambda,\Lambda^+_c$ and $\Lambda_b$
in $^{41}_j$Ca ($j=\Lambda,\Lambda^+_c,\Lambda_b$).
}
\label{Capot}
}
\hfill
\parbox{\halftext}{
\includegraphics[width=6.0cm,height=6.0cm,angle=-90]{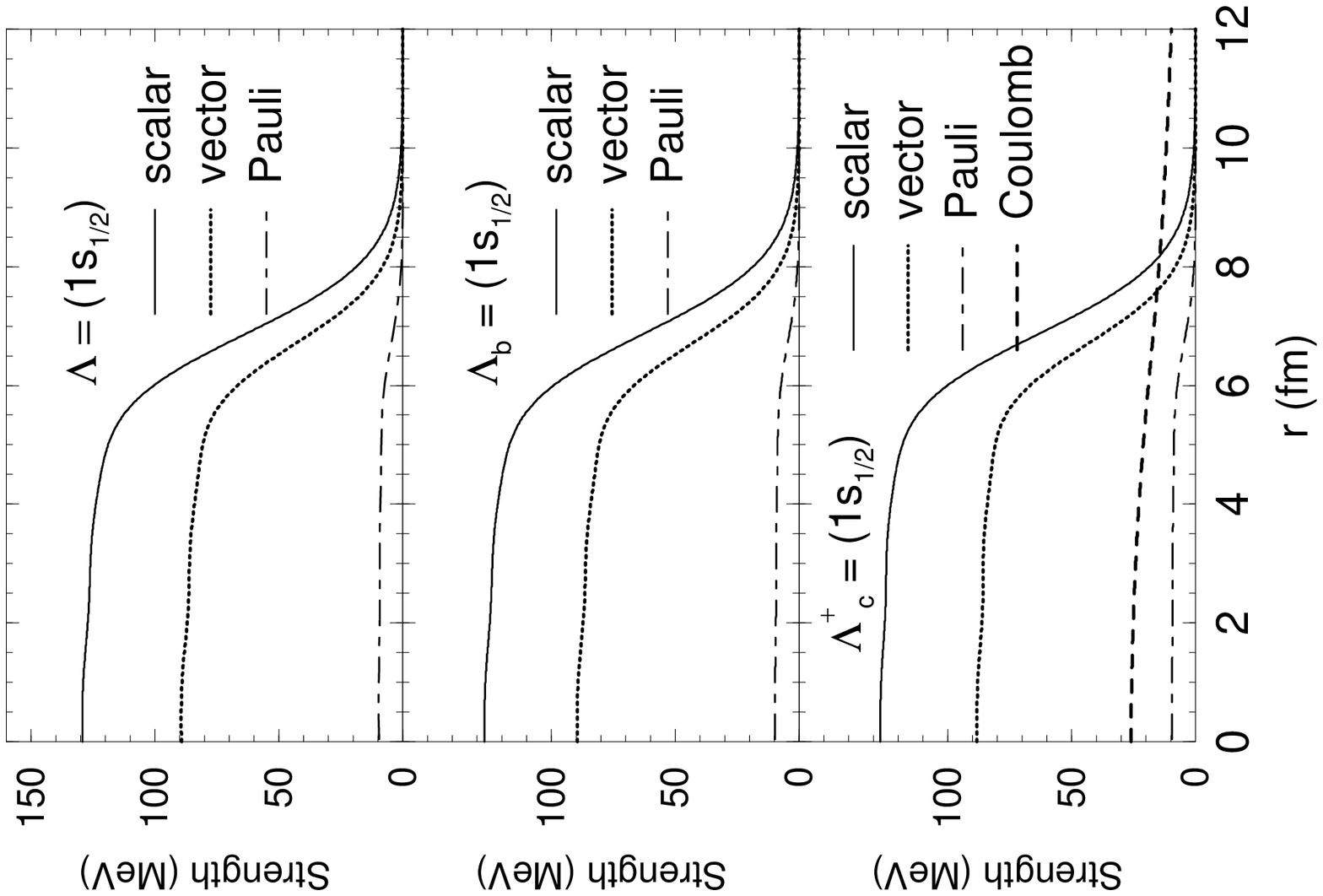}
\caption{Potential strengths for $1s_{1/2}$ state for 
the $\Lambda,\Lambda^+_c$ and $\Lambda_b$
in $^{209}_j$Pb ($j=\Lambda,\Lambda^+_c,\Lambda_b$).
}
\label{Pbpot}
}
\end{figure}

\begin{wrapfigure}{l}{6.0cm}
\includegraphics[width=6.0cm,height=6.0cm,angle=-90]{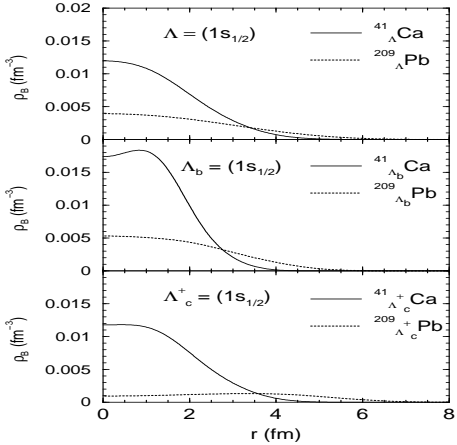}
\caption{$\Lambda,\Lambda^+_c$ and $\Lambda_b$
probability (baryon) density distributions for $1s_{1/2}$ state
in $^{41}_j$Ca and
$^{209}_j$Pb ($j = \Lambda,\Lambda^+_c,\Lambda_b$).
}
\label{prodensity}
\end{wrapfigure}

Next, in Figs.~\ref{Capot} and~\ref{Pbpot}, we show the scalar and vector
potential strengths for the $\Lambda$, $\Lambda^+_c$ and $\Lambda_b$
for $1s_{1/2}$ state
in $^{41}_j$Ca and $^{209}_j$Pb ($j=\Lambda,\Lambda^+_c,\Lambda_b$),
and the corresponding probability density distributions in 
Fig.~\ref{prodensity}.\hfill 
In Figs.~\ref{Capot} and~\ref{Pbpot}, 
"Pauli" stands for the effective, repulsive potential representing
the Pauli blocking at the quark level, plus
the $\Sigma_{c,b} N - \Lambda_{c,b} N$ channel coupling,
introduced at the hadronic level
phenomenologically~\cite{Tsushima_hyp}.
For the $\Lambda^+_c$, the Coulomb potentials are also shown.
The scalar and vector
potentials for these particles in hypernuclei multiplet
with the same baryon numbers are quite similar.
Thus, as far as the total baryon density distributions and
the scalar and vector potentials are concerned,
$\Lambda$, $\Lambda^+_c$ and $\Lambda_b$ hypernuclei show quite similar
features within the multiplet.

However, as shown in Fig.~\ref{prodensity},
the wave functions obtained for $1s_{1/2}$ state are very different.
The $\Lambda^+_c$ probability density distribution in $^{209}_{\Lambda^+_c}$Pb
is much more pushed away from the center
than that for the $\Lambda$ in $^{209}_\Lambda$Pb due to the Coulomb force.
On the contrary, the $\Lambda_b$ probability density distributions
in $\Lambda_b$ hypernuclei are much larger near the origin than those for
the $\Lambda$ in the corresponding $\Lambda$ hypernuclei
due to its heavy mass.

Finally, we show the calculated single-particle energies in
Tables~\ref{table1} and~\ref{table2}.
Results for the $\Lambda$ hypernuclei are from Ref.~\cite{Tsushima_hyp}.
\begin{table}[htbp]
\begin{center}
\caption{Single-particle energies (in MeV)
for $^{17}_j$O, $^{41}_j$Ca and $^{49}_j$Ca
($j=\Lambda,\Lambda^+_c,\Lambda_b$).
Results for the hypernuclei are taken  
from Ref.~\protect\cite{Tsushima_hyp}.
Spin-orbit splittings for $\Lambda$ hypernuclei 
are not well determined by the experiments.}
\label{table1}
\begin{tabular}[htbp]{c|cccc|cccc|ccc}
\hline 
&$^{16}_\Lambda$O  &$^{17}_\Lambda$O
&$^{17}_{\Lambda^+_c}$O  &$^{17}_{\Lambda_b}$O
&$^{40}_\Lambda$Ca &$^{41}_\Lambda$Ca
&$^{41}_{\Lambda^+_c}$Ca &$^{41}_{\Lambda_b}$Ca
&$^{49}_\Lambda$Ca    &$^{49}_{\Lambda^+_c}$Ca &$^{49}_{\Lambda_b}$Ca\\
&(Exp.~\protect\cite{chr})& & & &(Exp.~\protect\cite{chr})& & & & & & \\
\hline
$1s_{1/2}$&-12.5&-14.1&-12.8&-19.6&-20.0&-19.5&-12.8&-23.0&-21.0&-14.3&-24.4\\
$1p_{3/2}$&-2.5 &-5.1 &-7.3 &-16.5&-12.0&-12.3&-9.2 &-20.9&-13.9&-10.6&-22.2\\
$1p_{1/2}$&($1p_{3/2}$)&-5.0&-7.3 &-16.5
&($1p_{3/2}$)&-12.3&-9.1&-20.9&-13.8&-10.6&-22.2\\
$1d_{5/2}$&     &     &     &     &     &-4.7 &-4.8 &-18.4&-6.5 &-6.5 &-19.5\\
$2s_{1/2}$&     &     &     &     &     &-3.5 &-3.4 &-17.4&-5.4 &-5.3 &-18.8\\
$1d_{3/2}$&     &     &     &     &     &-4.6 &-4.8 &-18.4&-6.4 &-6.4 &-19.5\\
$1f_{7/2}$&     &     &     &     &     &     &     &     &---  &-2.0 &-16.8\\
\end{tabular}
\end{center}
\end{table}
%
%
\begin{table}[htbp]
\begin{center}
\caption{Single-particle energies (in MeV)
for $^{91}_j$Zr and $^{208}_j$Pb
($j=\Lambda,\Lambda^+_c,\Lambda_b$).
}
\label{table2}
\begin{tabular}[t]{c|cccc|cccc}
\hline
&$^{89}_\Lambda$Yb          &$^{91}_{\Lambda}$Zr
&$^{91}_{\Lambda^+_c}$Zr    &$^{91}_{\Lambda_b}$Zr
&$^{208}_\Lambda$Pb         &$^{209}_{\Lambda}$Pb
&$^{209}_{\Lambda^+_c}$Pb   &$^{209}_{\Lambda_b}$Pb\\
&(Exp.~\protect\cite{aji})& & & &(Exp.~\protect\cite{aji})& & & \\
\hline
$1s_{1/2}$&-22.5&-23.9&-10.8&-25.7&-27.0&-27.0&-5.2 &-27.4\\
$1p_{3/2}$&-16.0&-18.4&-8.7 &-24.2&-22.0&-23.4&-4.1 &-26.6\\
$1p_{1/2}$&($1p_{3/2}$)&-18.4&-8.7 &-24.2
&($1p_{3/2}$)&-23.4&-4.0 &-26.6\\
$1d_{5/2}$&-9.0 &-12.3&-5.8 &-22.4&-17.0&-19.1&-2.4 &-25.4\\
$2s_{1/2}$&---  &-10.8&-3.9 &-21.6&---  &-17.6&---  &-24.7\\
$1d_{3/2}$&($1d_{5/2}$)&-12.3&-5.8 &-22.4
&($1d_{5/2}$)&-19.1&-2.4 &-25.4\\
$1f_{7/2}$&-2.0 &-5.9 &-2.4 &-20.4&-12.0&-14.4&---  &-24.1\\
$2p_{3/2}$&---  &-4.2 &---  &-19.5&---  &-12.4&---  &-23.2\\
$1f_{5/2}$&($1f_{7/2}$)&-5.8 &-2.4 &-20.4
&($1f_{7/2}$)&-14.3&---  &-24.1\\
$2p_{1/2}$&     &-4.1 &---  &-19.5&---  &-12.4&---  &-23.2\\
$1g_{9/2}$&     &---  &---  &-18.1&-7.0 &-9.3 &---  &-22.6\\
$1g_{7/2}$&     &     &     &
&($1g_{9/2}$)&-9.2 &---  &-22.6\\
$1h_{11/2}$&    &     &     &     &     &-3.9 &---  &-21.0\\
$2d_{5/2}$&     &     &     &     &     &-7.0 &---  &-21.7\\
$2d_{3/2}$&     &     &     &     &     &-7.0 &---  &-21.7\\
$1h_{9/2}$&     &     &     &     &     &-3.8 &---  &-21.0\\
$3s_{1/2}$&     &     &     &     &     &-6.1 &---  &-21.3\\
$2f_{7/2}$&     &     &     &     &     &-1.7 &---  &-20.1\\
$3p_{3/2}$&     &     &     &     &     &-1.0 &---  &-19.6\\
$2f_{5/2}$&     &     &     &     &     &-1.7 &---  &-20.1\\
$3p_{1/2}$&     &     &     &     &     &-1.0 &---  &-19.6\\
$1i_{13/2}$&    &     &     &     &     &---  &---  &-19.3\\
\end{tabular}
\end{center}
\end{table}
%
%
Recall that since the mass difference for $\Lambda_c^+$ and $\Sigma_c$, 
and probably for $\Lambda_b$ and $\Sigma_b$, are larger
than that for $\Lambda$ and $\Sigma$,
we expect the effect of the channel coupling for the charmed and bottom 
hypernuclei to be smaller than those for the
strange hypernuclei, although the same parameters are used.
In addition, we searched for the single-particle states up to
the same highest state as that of the core neutrons in each hypernucleus,
since the deeper levels are usually easier to observe in experiment.

First, it is clear that the $\Lambda^+_c$
single-particle energy levels are higher
than the corresponding levels for the $\Lambda$ and $\Lambda_b$.
This is because of the
Coulomb force. This feature becomes stronger as
the proton number increases.

Second, the level spacing for the $\Lambda_b$ single-particle energies
is much smaller than that for the $\Lambda$ and $\Lambda^+_c$.
This is due to the heavy mass of $\Lambda_b$ (or $M^\star_b$).
In the Dirac equation for the $\Lambda_b$,
the mass term dominates more than that of the term
proportional to Dirac's $\kappa$, which classifies the states,
or single-particle
wave functions. (See Refs.~\cite{Saitof,Tsushima_hyp} for detail.)
This small level spacing would make it very difficult to
distinguish the states in experiment,
or to achieve such high resolution.
On the other hand, this may imply also many new phenomena.
It will have a large probability to trap a $\Lambda_b$ among
one of those many states,
especially in heavy nucleus.
What are the consequences ? May be the
$\Lambda_b$ weak decay happens inside a heavy nucleus with a very low
probability ? Does it emit many photons when the $\Lambda_b$ gradually
makes transitions from a deeper state to a shallower state ?
All these questions raise a flood of speculations.

Finally, it should be emphasized that we have used the values
for the coupling constants of $\sigma$
(or $\sigma$-field dependent strength), $\omega$ and $\rho$
to $\Lambda, \Lambda^+_c$ and $\Lambda_b$ to be determined
automatically based on the underlying quark model, as for the
nucleon and other baryons.
(Recall that the values for the vector $\omega$ fields to
any baryons can be obtained by the number of light quarks in a baryon,
but those for the $\sigma$ are different as
shown in Eqs.~(\ref{gsigma})~-~(\ref{bagradii}).)
Phenomenology would determine ultimately if the coupling constants
(strengths) determined by the underlying quark model
actually work for $\Lambda^+_c$ and $\Lambda_b$ or not.
Although implications of the present results can be speculated
a great deal, we would like to emphasize that,
what we showed is that the $\Lambda^+_c$ and $\Lambda_b$ hypernuclei
would exist in realistic experimental conditions.
Experiments at facilities like JHF would provide further inputs to gain
a better understanding of the interaction of $\Lambda^+_c$
and $\Lambda_b$ with the nuclear matter.
Additional studies are needed to investigate the semi-leptonic
weak decay of $\Lambda^+_c$ and $\Lambda_b$.
The role of Pauli blocking and density in influencing the decay rates
as compared to those for the free hyperons would be highly useful.
Will the high density lead to a slower decay and that a higher
probability to survive its passage through the material ?
At present the study of the presence of
$\Lambda^+_c$ and $\Lambda_b$ in finite nuclei is its infancy.
Careful investigations, both theoretical and experimental, would lead to
a much better understanding of the role of heavy quarks in finite nuclei, 
and the role of partial restoration of chiral symmetry in nuclear medium.

\section*{Acknowledgments}
K.T. would like to thank the organizers of the workshop, 
especially Profs. T. Kunihiro and A. Hosaka for the support. 
Our thanks go to Prof. A.W. Thomas for the hospitality
at CSSM, Adelaide, where this work was initiated.
K.T. acknowledges support and warm hospitality at University of
Alberta, where most of the work reported here was completed.
K.T. is supported by the Forschungszentrum-J\"{u}lich,
contract No. 41445282 (COSY-058). The work of F.K. is
supported by NSERCC.

%

\end{document}